\documentclass[runningheads]{llncs}
\usepackage[T1]{fontenc}
\usepackage{graphicx,verbatim}
\usepackage{amsmath}
\usepackage{amsfonts}
\usepackage{multirow}
\usepackage{booktabs}
\usepackage{hyperref}
\usepackage{pifont}

\begin{document}
\title{SWinMamba: Serpentine Window State Space Model for Vascular Segmentation}

\author{Rongchang Zhao\inst{1} \and
Huanchi Liu\inst{1}\and
Jian Zhang\inst{1}}

\authorrunning{R.Zhao et al.}
%
\institute{School of Computer Science, Central South University
\email{zhaorc@csu.edu.cn}}
\maketitle              

\begin{abstract}
Vascular segmentation in medical images is crucial for disease diagnosis and surgical navigation. However, the segmented vascular structure is often discontinuous due to its slender nature and inadequate prior modeling. In this paper, we propose a novel Serpentine Window Mamba (SWinMamba) to achieve accurate vascular segmentation. The proposed SWinMamba innovatively models the continuity of slender vascular structures by incorporating serpentine window sequences into bidirectional state space models. The serpentine window sequences enable efficient feature capturing by adaptively guiding global visual context modeling to the vascular structure. Specifically, the Serpentine Window Tokenizer (SWToken) adaptively splits the input image using overlapping serpentine window sequences, enabling flexible receptive fields (RFs) for vascular structure modeling. The Bidirectional Aggregation Module (BAM) integrates coherent local features in the RFs for vascular continuity representation. In addition, dual-domain learning with Spatial-Frequency Fusion Unit (SFFU) is designed to enhance the feature representation of vascular structure. Extensive experiments on three challenging datasets demonstrate that the proposed SWinMamba achieves superior performance with complete and connected vessels. 
\end{abstract}

\section{Introduction}
Vascular segmentation is crucial in medical image analysis, substantially benefiting disease diagnosis and treatment. The topological structure of the vascular system serves as a critical biomarker for assessing cardiovascular disease, diabetic retinopathy, and tumor microenvironments~\cite{DR,microenvironment,cardiovascular}. Obtaining complete and connected vascular structures provides clinicians with quantitative parameters such as vascular diameter, curvature, and bifurcation angles, which are essential for disease diagnosis and surgical navigation~\cite{param_stroke,param_planning,param_navigation}.

\begin{figure}[t!]
\includegraphics[width=\textwidth]{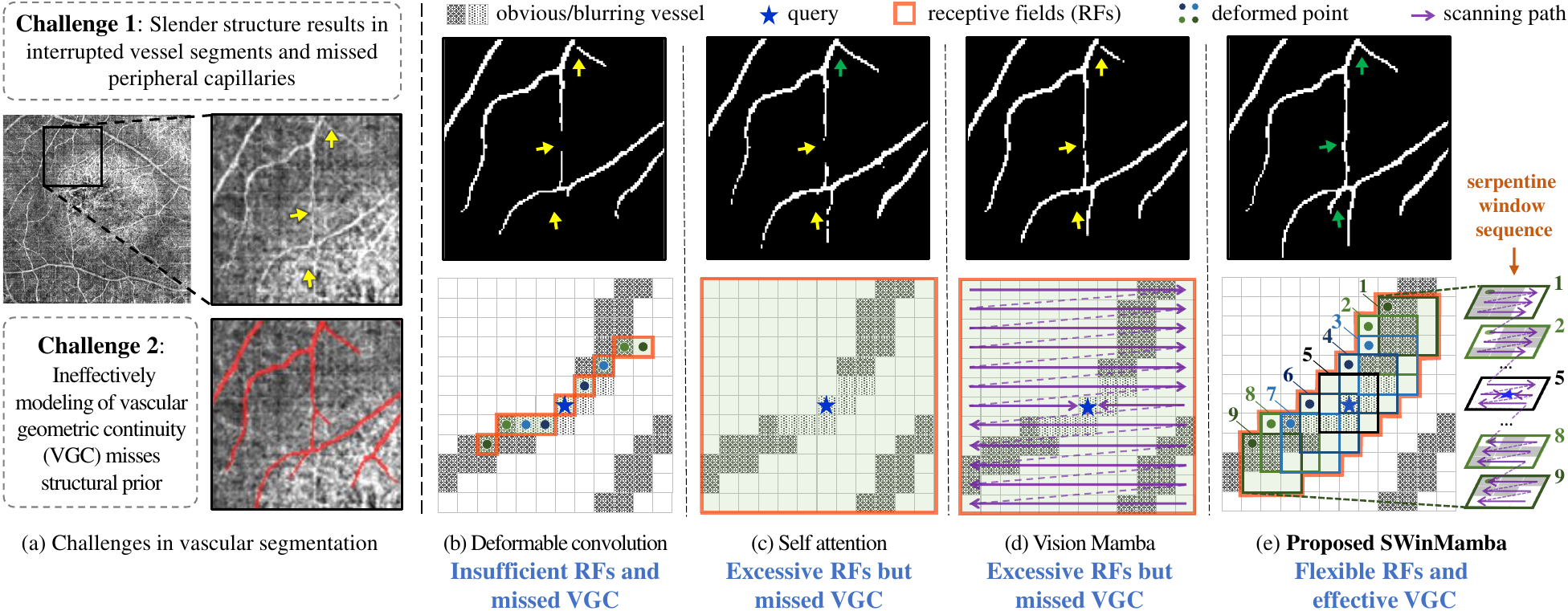}
\caption{\textbf{Challenges.} Vascular structures exhibit slender and fragile morphology, posing significant challenges for segmentation algorithms. \textbf{Motivation.} The proposed SWinMamba adaptively guides global visual context modeling to the serpentine window sequences to efficiently model the continuity of slender vascular structures with the bidirectional state space models.} \label{fig1}
\end{figure}

However, accurately segmenting vascular structures in medical images is challenging due to their complex morphology (Fig.~\ref{fig1}(a)). On the one hand, the slender nature of vessels makes them susceptible to background noise and low contrast, resulting in interrupted vascular segments and missed peripheral capillaries.
On the other hand, effectively modeling the intrinsic \textit{Vascular Geometric Continuity (VGC)} remains an open challenge. VGC refers to the continuous variations in diameter and direction along vascular structures~\cite{VGC1989}, providing crucial structural priors for accurately annotating blurry segments. This limitation manifests particularly in complex vascular networks where gradual tapering and bifurcation patterns contain critical diagnostic information. 

Despite numerous efforts, existing approaches still fail to obtain complete and connected vessels. 
Initially, convolution-based methods~\cite{MDANet,EDAE-Net,DSCNet,UNet} (Fig.~\ref{fig1}(b)) detect vessel structures using standard or modified kernels that integrate strip priors, deformation, or dilation operations. However, these methods struggle to capture the continuous structures of vascular diameter and direction due to limited receptive fields (RFs). Recently, hybrid methods~\cite{TransUNet,DE-DCGCN-EE,GT-DLA-dsHFF,CS2Net} (Fig.~\ref{fig1}(c)) have been proposed to expand RFs using self-attention or graph technologies. Nevertheless, these approaches still face challenges such as heavy computational burdens and difficulty in effectively focusing on slender vessels. Moreover, existing methods model vascular structure as a set of isolated pixels rather than a series of interconnected tubular segments, further limiting their ability to extract sequential dependencies, including VGC, between adjacent segments.

The emerging state space model (SSM), exemplified by Mamba~\cite{Mamba}, presents a compelling framework for modeling sequential dependencies among tubular segments in vascular structure. This potential stems from Mamba's inherent capacity to dynamically capture long-range contextual patterns through efficient selective and ordered processing. However, Vision Mamba (Vim, Fig.~\ref{fig1}(d))~\cite{Vision_Mamba} in image segmentation faces obstacles: \textbf{1) Inefficient tokenization} disrupts spatial coherence by flattening 2D/3D vascular structures into rigid 1D sequences, decoupling locally coherent features and introducing spurious connectivity in bifurcation regions. \textbf{2) Rigid scanning} ignores vascular topology, obscuring geometric gradients through inconsistent sampling and confusing the model under incomplete visual conditions. \textbf{3) Failed VGC modeling.} The sequential inductive bias of Vim conflicts with the vascular geometrical priors of history-dependent morphological momentum and non-local curvature constraints. 
\begin{figure}[t!]
\includegraphics[width=\textwidth]{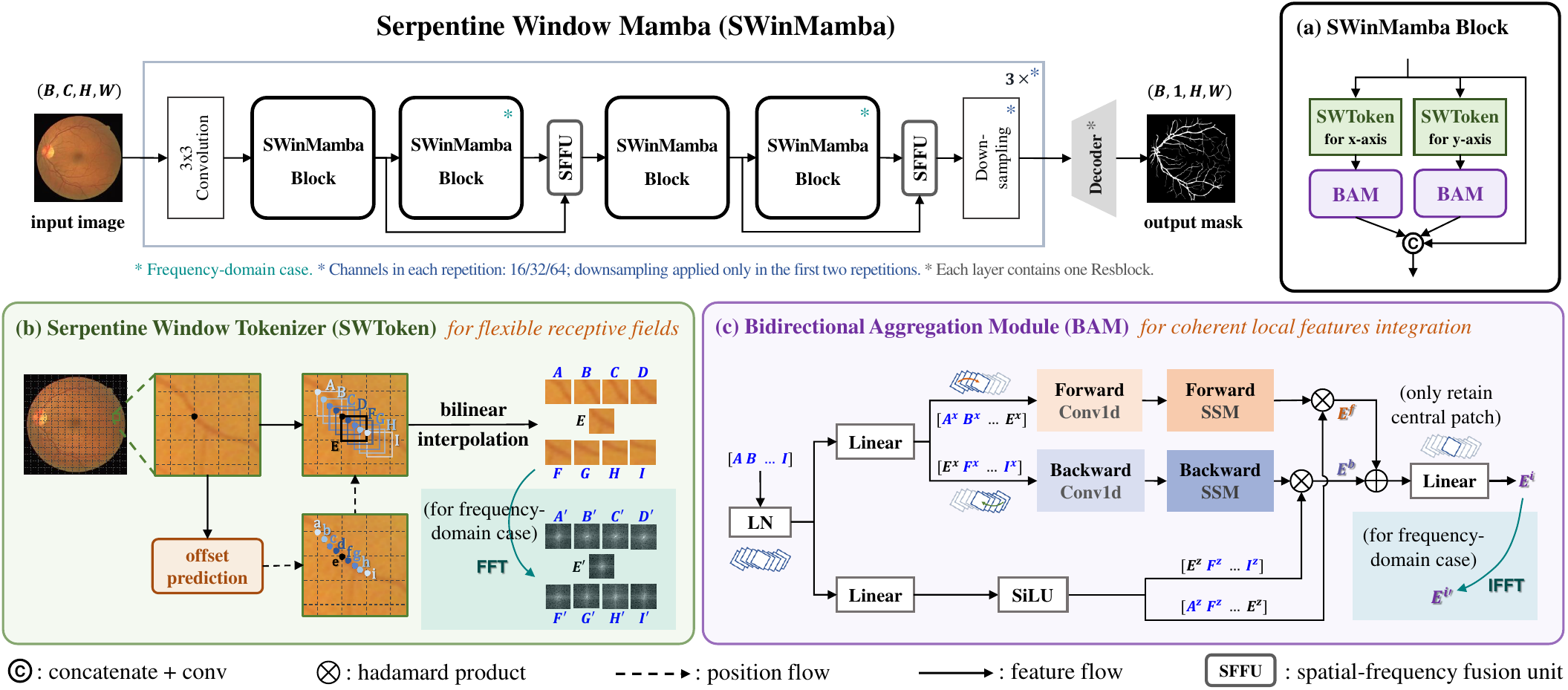}
\caption{SWinMamba Overview. The proposed SWinMamba adopts a U-shaped structure where the encoder includes the novel SWinMamba block and feature fusion unit (SFFU). The SWinMamba block incorporates serpentine window sequences to bidirectional state space models to guide the global context modeling to the continuous vascular structure. The SWinMamba block includes a newly designed tokenizer (SWToken) for flexible receptive fields and bidirectional aggregation module (BAM) for the integration of coherent local features.} 
\label{fig2}
\end{figure}
In this paper, we propose a novel Serpentine Window Mamba (SWinMamba, Fig.\ref{fig2}) to accurately segment vascular structures from medical images. The proposed SWinMamba ensures complete and connected segmentation results by modeling the continuity of slender vessels as sequential representations of serpentine window contents, relying on three innovative components: \textbf{1) Serpentine Window Tokenizer (SWToken)} to provide flexible and learnable RFs for vascular representation; \textbf{2) Bidirectional Aggregation Module (BAM)} to integrate coherent local features, especially VGC, by sequentially scanning serpentine windows along vascular structures; and \textbf{3) the Spatial-Frequency Fusion Unit (SFFU)} to enable comprehensive vascular representation for accurate segmentation.

Our main contributions are threefold: 1) For the first time, a novel framework (SWinMamba) is proposed to achieve accurate vascular segmentation by incorporating serpentine window sequences into bidirectional state space models. 
2) A novel tokenization strategy (SWToken) is designed to adaptively split the input image using overlapping windows in a serpentine manner (serpentine window sequences), providing flexible receptive fields for vascular structure modeling. 
3) An ordered modeling approach (BAM) is proposed to integrate coherent local features into vascular continuity representation by modeling the dependencies of serpentine window sequences.

\section{Method} 
The Serpentine Window Mamba (SWinMamba, Fig.~\ref{fig2}) is proposed to achieve complete and connected vascular segmentation through a novel sequential modeling approach. It consists of 1) the SWinMamba block (Fig.~\ref{fig2}(a)), which sequentially traverses serpentine windows to model vascular continuity along local vessel structures, including the Serpentine Window Tokenizer (SWToken) for flexible and learnable receptive fields and the Bidirectional Aggregation Module (BAM) for sequential integration of coherent local features, and 2) the Spatial-Frequency Fusion Unit (SFFU) for comprehensive feature representation, enabling complete and connected vascular segmentation.

\subsection{Serpentine Window Tokenizer (SWToken) for flexible RFs}
The SWToken (Fig.~\ref{fig2}(b)) adaptively splits the input image using overlapping window sequences in a serpentine manner, providing efficient receptive fields to capture vascular continuity.

\noindent\textbf{(1) Tortuous Structure Locating.} To accurately focus receptive fields on the elongated and tortuous vascular structures, multiple strings of serpentine anchor points, adaptive to vessel morphology, are first located to anchor local segments. Given the input feature map $F\in\mathbb R^{C\times H\times W}$, interval $s$ that divides both $H$ and $W$, and string length $L$, the coordinates of the central anchor points for each string can be directly calculated as follows:
\begin{equation}
(x^{h,w}_c, y^{h,w}_c)=(h\times s, w\times s), \;h\in\{0,1,...,\frac{H}{s}-1\}, w \in\{0,1,...,\frac{W}{s}-1\},
\label{eq1_fixed-anchor}
\end{equation}
where $c=\lfloor\frac{L}{2}\rfloor$. The remaining anchor points adaptively extend along the x-axis or y-axis to align with vessel contours. Their coordinates are dynamically determined via a cumulative process, based on learnable offsets $\Delta \in [-1,1]$ that obtained from applying once convolution with a stride of $s$ to $F$. Eq.~\ref{eq2_dynamic-anchor} describes the computation for extension along the x-axis, with the y-axis similarly.
\begin{equation}
\left\{
\begin{aligned}
(x_{c+i}^{h,w},y_{c+i}^{h,w})\!&=\!(x^{h,w}_c+\alpha i,y^{h,w}_c+\!\sum\nolimits^{c+i}_{k=c}\!\alpha\Delta_{ y_k}), \\
(x_{c-i}^{h,w},y_{c-i}^{h,w})\!&=\!(x^{h,w}_c-\alpha i,y^{h,w}_c+\!\sum\nolimits^c_{k=c-i}\!\alpha\Delta_{y_k}),
\end{aligned}
\right.
\label{eq2_dynamic-anchor}
\end{equation}
where $\alpha$ is the fixed extend stride, and $i\in\{1,2,...,\lfloor\frac{L}{2}\rfloor\}$. In this paper, $L=9$, $s=8$, and $\alpha=2$ are set empirically.

\noindent\textbf{(2) Geometric Information Capturing.} To ensure receptive fields cover the complete vessel diameter and direction for vascular representation, tokens within $s\times s$ windows, which positioned by anchor points, are densely sampled from $F$ to capture sufficient geometric information. The coordinates within each window are determined by adding a relative offset to that of the corresponding anchor point. 
Since these coordinates are often fractional, the sampling is implemented via bilinear interpolation and can be represented as follows:
\begin{equation}
t^{h,w}_{l,m,n} = B(loc^{h,w}_{l,m,n}), \;loc^{h,w}_{l,m,n}=(x^{h,w}_l+m, y^{h,w}_l+n), \;T^{h,w}_{l,m,n} \in \mathbb R^C
\label{eq3_token}
\end{equation}
where $B(\cdot)$ represents bilinear interpolation, corresponding anchor point index $l\in\{0,1,...,L-1\}$, and relative offset $m,n\in\{0,1,...,s-1\}$. The input to SWToken is a feature map of size $(C,H,W)$, and the output consists of $\frac{HW}{s^2}$ token sets, each containing $Ls^2$ tokens from $L$  windows.

\noindent\textbf{Summary of Advantages:} The SWToken splits the input image using overlapping windows to offer flexible receptive fields for efficient vascular representation construction. 

\subsection{Bidirectional Aggregation Module (BAM) for coherent local features integration}
BAM (Fig.~\ref{fig2}(c)) integrates coherent local features to represent sequential dependencies in vascular structure, especially VGC, by incorporating a bidirectional aggregation mechanism into the state space model.
BAM processes the input $\frac{HW}{s^2}$ sets of tokens generated by SWToken in the same manner: Tokens within each window are first arranged into sequences as follows:
\begin{equation}
T^{h,w}_l=[t^{h,w}_{l,0,0},t^{h,w}_{l,0,1},...,t^{h,w}_{l,1,0},t^{h,w}_{l,1,1},...,t^{h,w}_{l,s-1,s-1}], \;T^{h,w}_l\in\mathbb R^{C\times s^2}.
\label{eq4_patch}
\end{equation}
The bidirectional aggregation procedure, which leverages Mamba to extract local features during the scanning of individual sequences and integrate their continuity through cross-sequence traversal, can be represented as:
\begin{equation}
\begin{aligned}
S^{h,w}_{out}=\psi(Mamba_f(S^{h,w}_f))+\varphi(\psi(&Mamba_b(\varphi(S^{h,w}_b)))), \;S^{h,w}_{out}\in\mathbb R^{C\times s^2},\\ 
S^{h,w}_f=[T^{h,w}_0,T^{h,w}_1,...,T^{h,w}_c&],\;S^{h,w}_b=[T^{h,w}_c, T^{h,w}_{c+1},...,T^{h,w}_{L-1}],
\end{aligned}
\label{eq5_BCM}
\end{equation}
where $S^{h,w}_{out}$ denotes the processed sequence, $\varphi(\cdot)$ denotes the reversal operation, and $\psi(\cdot)$ extracts the last $s^2$ elements of the sequence. The final output of BAM is a feature map of shape $(C,H,W)$, formed by reshaping each of the $\frac{HW}{s^2}$ processed sequences into an $s\times s$ patch and concatenating them according to superscripts.

\noindent\textbf{Summary of Advantages:} The BAM enables SWinMamba to integrate coherent local features into the representation of vascular continuity. 

\subsection{Spatial-Frequency Fusion Unit (SFFU) for comprehensive feature representation}

The SFFU integrates vascular features from the spatial and frequency domains to construct comprehensive representations for complete and connected vascular segmentation. 
In frequency-domain case, contents of each window are transformed via Fast Fourier Transform (FFT), with their real and imaginary parts stacked, before being fed into the BAM. Correspondingly, patches reshaped from processed sequences undergo inverse FFT before being concatenated. The SFFU is represented as:
\begin{equation}
\begin{aligned}
\mathcal F_{fuse}={attn\odot \mathcal F_{spa}}+{(1-attn)\odot \mathcal F_{fre}},\, attn=SA(\mathcal F_{spa})\odot CA(\mathcal F_{spa}),\label{eq4}
\end{aligned}
\end{equation}
where $\odot$ denotes the Hadamard product; $\mathcal F_{spa}$ and $\mathcal F_{fre}$ represent the spatial-domain and frequency-domain features, respectively. The spatial attention $SA(\cdot)$ and channel attention $CA(\cdot)$ share the same computation as~\cite{CBAM} and are applied to $s\times s$ non-overlapping patches, rather than the entire image.

\noindent\textbf{Summary of Advantages:} SFFU empowers SWinMamba with comprehensive feature representation for complete and connected vascular segmentation.

\section{Experiments and Results Analysis}
\subsection{Datasets}
Three challenging datasets with varying modalities and vascular morphologies, CHASE-DB1~\cite{CHASE-DB1}, OCTA-500~\cite{OCTA-500}, and DCA1~\cite{DCA1}, are employed to evaluate the effectiveness of SWinMamba. The CHASE-DB1 dataset has 28 color fundus of the size $999\times 960$. The OCTA-500 dataset is a large-scale benchmark for Optical Coherence Tomography Angiography (OCTA) images, and its sub-dataset 6M employed here contains 300 samples with a resolution of $400\times 400$. The DCA1 dataset consists of 134 X-ray coronary angiography images, each sized $300\times300$.

\subsection{Implementation Details and Evaluation Metrics}
  
Data augmentation, including translation, rotation, mirroring, flipping, and random cropping with a fixed size of $256\times 256$, is performed simultaneously to avoid over-fitting.
The Adam optimizer with an initial learning rate of 1e-4 is utilized to optimize learnable parameters. The total epoch is set to 800, and the batch size is set to 1 due to hardware limitations. Four-fold cross-validation is adopted. 

For a fair quantitative comparison with other methods, SWinMamba is evaluated with three metrics: centerline Dice (clDice)~\cite{clDice}, the Betti error for Betti number $\beta_0$, and Dice coefficient (Dice), which measure completeness, connectivity, and overall accuracy, respectively. In addition, floating point operations per second (FLOPs) and number of parameters (Params) are utilized to assess the computational complexity of each model. 

\subsection{Evaluation Performance on Three Challenging Datasets}
\begin{table}[t] 
    \centering
    \caption{Experimental results demonstrate that the proposed SWinMamba successfully achieves vascular segmentation in different medical images with the best results.}
    \resizebox{1\textwidth}{!}{      
    \begin{tabular}{cccccccccccccc}
         \toprule
         \multicolumn{2}{c}{\multirow{2.5}{*}{Method}} & \multirow{2.5}{*}{Year} & \multicolumn{3}{c}{\multirow{1.4}{*}{~~~CHASE-DB1}} & \multicolumn{3}{c}{\multirow{1.4}{*}{~~~OCTA-500}} & \multicolumn{3}{c}{\multirow{1.4}{*}{~~~DCA1}} & \raisebox{-3pt}{FLOPs} & \raisebox{-3pt}{Params}\\
         \cmidrule(lr{0pt}){4-6}\cmidrule(lr{0pt}){7-9}\cmidrule(lr{0pt}){10-12}
         \multicolumn{2}{c}{} &  & ~~clDice$\uparrow$ & $\beta_0\downarrow$ & Dice$\uparrow$ & ~~clDice$\uparrow$ & $\beta_0\downarrow$ & Dice$\uparrow$ & ~~clDice$\uparrow$ & $\beta_0\downarrow$ & Dice$\uparrow$ & \raisebox{2pt}{(G)$\downarrow$} & \raisebox{2pt}{(M)$\downarrow$} \\
         \toprule
         \multicolumn{2}{c}{UNet} & 2015 & ~~.7817 & 2.252 & .7880 & ~~.8674 & 3.603 & .8353 & ~~.7990 & 1.526 & .7445 & ~~40.05 & 17.26\\
         \multicolumn{2}{c}{MDANet} & 2022 & ~~.8241 & 1.891 & .8176 & ~~.8696 & 3.140 & .8352 & ~~.8245 & 1.470 & .7643 & ~~33.37 & 13.83\\
         \multicolumn{2}{c}{DSCNet} & 2023 & ~~.8264 & \textbf{1.434} & .8215 & ~~.8624 & 3.639 & .8291 & ~~.8176 & 1.434 & .7606 & ~~7.67 & \textbf{2.09}\\
         \multicolumn{2}{c}{EDAE-Net} & 2024 & ~~.8292 & 1.675 & .8213 & ~~.8722 & 3.042 & .8382 & ~~.8278 & 1.705 & .7638 & ~~\textbf{3.99} & 35.54\\
         \midrule
         \multicolumn{2}{c}{CS$^2$-Net} & 2020 & ~~.8191 & 1.556 & .8151 & ~~.8681 & 3.875 &.8348 & ~~.8186 & 1.600 & .7621 & ~~32.00 & 8.92\\
         \multicolumn{2}{c}{TransUNet} & 2021 & ~~.8016 & 1.734 & .8008 & ~~.8681 & 3.332 & .8330 & ~~.8212 & 1.717 & .7616 & ~~33.74 & 88.06\\
         \multicolumn{2}{c}{DE-DCGCN-EE} & 2022 & ~~.8236 & 2.020 & .8152 & ~~.8719 & 3.550 & .8378 & ~~.8256 & 1.487 & \textbf{.7660} & ~~295.03 & 14.11\\
         \multicolumn{2}{c}{GT-DLA-dsHFF} & 2023 & ~~.8236 & 1.939 & .8185 & ~~.8698 & 3.189 & .8396 & ~~.8232 & 1.717 & .7653 & ~~473.92 & 26.08\\
         \midrule
         \multicolumn{2}{c}{U-Mamba} & 2024 & ~~.8267 & 1.879 & .8202 & ~~.8618 & 3.407 & .8259 & ~~.8105 & 1.502 & .7601 & ~~54.65 & 46.39\\
         \multicolumn{2}{c}{VM-UNet} & 2024 & ~~.8204 & 2.734 & .8160 & ~~.8704 & 4.259 & .8346 & ~~.8234 & 1.452 & .7618 & ~~4.09 & 27.43\\
         \midrule
         \multicolumn{2}{c}{\textbf{SWinMamba}} & 2025 & ~~\textbf{.8373} & 1.546 & \textbf{.8243} & ~~\textbf{.8730} & \textbf{2.636} & \textbf{.8403} & ~~\textbf{.8286} & \textbf{1.378} & .7659 & ~~8.20 & 2.78\\    
         \bottomrule
    \end{tabular}
    }
    \label{table1_comparison}
\end{table}
Quantitative results (Table~\ref{table1_comparison}) and visualization (Fig.~\ref{fig3}) on three datasets demonstrate that the proposed SWinMamba consistently achieves superior performance with connected and complete vascular segmentation. 
\textbf{1) Results on CHASE-DB1} indicate SWinMamba obtains outstanding performance at different levels (diameters) of the vascular structure, with 0.8373 of clDice, 1.546 of $\beta_0$, and 0.8243 of Dice.
\textbf{2) Results on OCTA-500} demonstrate that SWinMamba achieves robust vascular segmentation, with clDice of 0.8730, $\beta_0$ of 2.636, and Dice of 0.8403. It should be noted that SWinMamba effectively segments the connected vessels from noisy and complex backgrounds by leveraging the structural prior from VGC. 
\textbf{3) Results on DCA1} show that SWinMamba delivers clDice of 0.8286, $\beta_0$ of 1.378, and Dice of 0.7659 on X-ray coronary angiography. It indicates that SWinMamba achieves connected and complete vascular segmentation for coronary angiography with sparse density of vascular structures. 
\subsection{Ablation Study}
\begin{table}[t]
    \centering
    \caption{Extensive ablation studies on three datasets present the effectiveness of each component of the SWinMamaba. The best results are marked in \textbf{bold}. F: Applying frequency-domain-case transformation.} 
    \resizebox{1.0\textwidth}{!}{        
    \begin{tabular}{cccccccccccccc}
         \toprule
         \multicolumn{1}{c}{\multirow{2.5}{*}{Method}} & \multicolumn{4}{c}{\multirow{1.4}{*}{~~Modules}} & \multicolumn{3}{c}{\multirow{1.4}{*}{~~~CHASE-DB1}} & \multicolumn{3}{c}{\multirow{1.4}{*}{~~~OCTA-500}} & \multicolumn{3}{c}{\multirow{1.4}{*}{~~~DCA1}}\\
         \cmidrule(lr{0pt}){2-5}\cmidrule(lr{0pt}){6-8}\cmidrule(lr{0pt}){9-11}\cmidrule(lr{0pt}){12-14} 
         {} & ~~BAM & SWToken & F & SFFU & ~~clDice$\uparrow$ & $\beta_0\downarrow$ & Dice$\uparrow$ & ~~clDice$\uparrow$ & $\beta_0\downarrow$ & Dice$\uparrow$ & ~~clDice$\uparrow$ & $\beta_0\downarrow$ & Dice$\uparrow$\\
         \toprule
         Baseline &  &  &  &  & ~~{.7920} & {2.647} & {.7885} & ~~{.8678} & {4.196} & {.8311} & ~~{.8021} & {2.818} & {.7518} \\
         M1 & ~~\checkmark &  &  &  & ~~.8247 & 1.743 & .8135 & ~~.8694 & 3.691 & .8366 & ~~.8137 & 1.738 & .7550 \\
         M2 & ~~\checkmark & \checkmark &  &  & ~~.8278 & 1.604 & .8176 & ~~.8712 & 2.706 & .8389 & ~~.8244 & 1.393 & .7627 \\
         M3 & ~~\checkmark & \checkmark & \checkmark &  & ~~.8267 & 1.832 & .8141 & ~~.8719 & 3.207 & .8378 & ~~.8170 & 1.474 & .7576 \\
         \textbf{SWinMamba} & ~~\checkmark & \checkmark & \checkmark & \checkmark & ~~\textbf{.8373} & \textbf{1.546} & \textbf{.8243} & ~~\textbf{.8730} & \textbf{2.636} & \textbf{.8403} & ~~\textbf{.8286} & \textbf{1.378} & \textbf{.7659}\\
         \bottomrule     
    \end{tabular}
    }
    \label{table2_ablation}
\end{table}

Extensive ablation experiments in Table~\ref{table2_ablation} are performed on three datasets to verify the contribution of each component. \noindent\textbf{1) Effectiveness of BAM.} Directly adding BAM (equivalent to adding Vim block~\cite{Vision_Mamba}) generally improves segmentation performance, demonstrating that bidirectional aggregation processing can extract effective vascular features. \noindent\textbf{2) Effectiveness of SWToken.} Comparing M1 with M2 shows that incorporating SWToken significantly enhances all evaluation metrics, achieving average improvements of 18.17\%, 0.63\%, and 0.60\%,on $\beta_0$, clDice and Dice, respectively. This illustrates that the flexible receptive fields provided by SWToken, which guides the BAM to focus on slender vessels and aligns its sequential computation with vascular structures, plays a critical role in constructing effective vascular representations. \noindent\textbf{3) Effectiveness of SFFU.} Adding FFT alone (M3) leads to a general decline in metrics (except for clDice on OCTA-500), while further incorporating SFFU results in significant improvements over M2, proving that SFFU facilitates the construction of comprehensive feature representations for complete vascular segmentation.

\subsection{Comparison with State-of-the-art (SOTA)}
\begin{figure}[t!]
\includegraphics[width=1.0\textwidth]{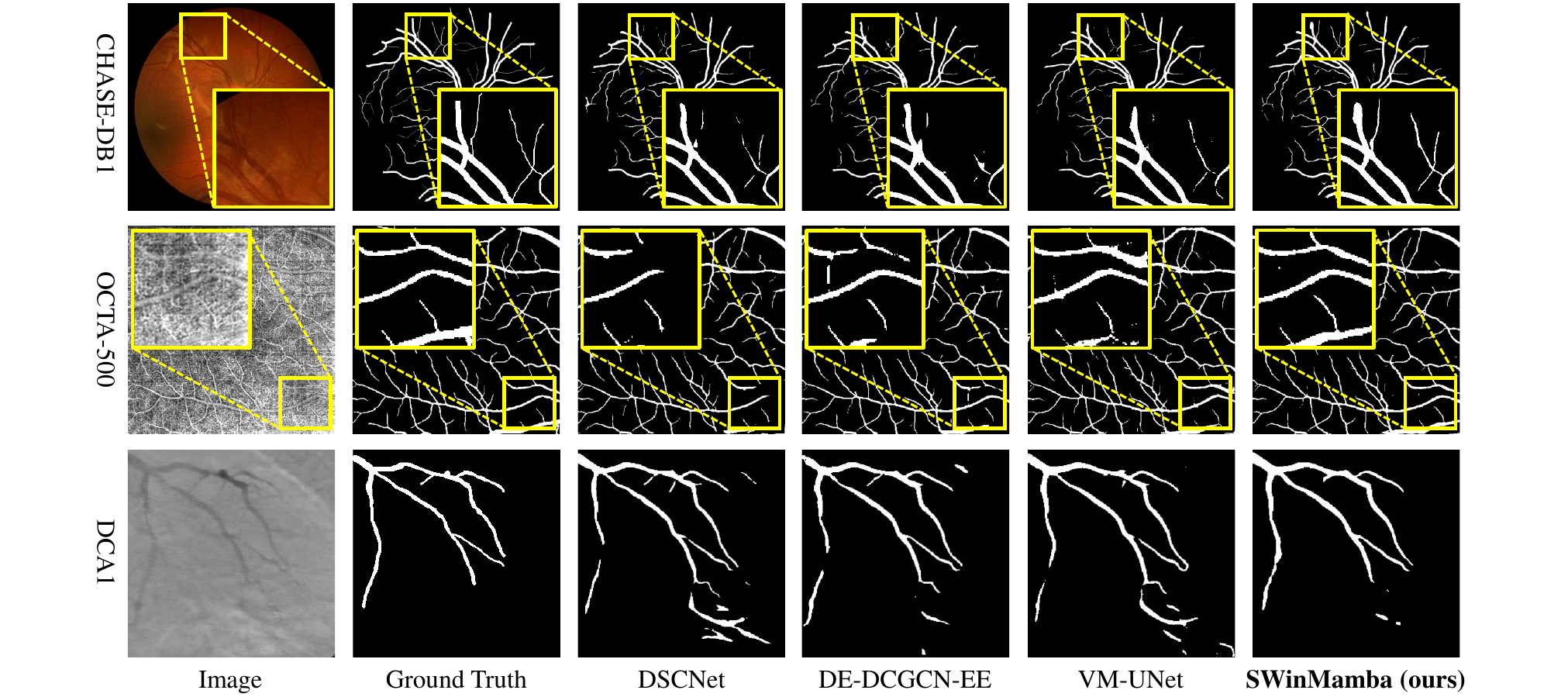}
\caption{Visualization shows that SWinMamba achieves complete and connected vascular segmentation across three modalities of medical images.} \label{fig3}
\end{figure}
Comparison experiments (Table~\ref{table1_comparison}) illustrate the superiority and generalizability of SWinMamba. The experiments include four convolution-based methods (UNet~\cite{UNet}, MDANet~\cite{MDANet}, DSCNet~\cite{DSCNet}, and EDAE-Net~\cite{EDAE-Net}), four hybrid methods (CS$^2$-Net~\cite{CS2Net}, TransUNet~\cite{TransUNet}, DE-DCGCN-EE~\cite{DE-DCGCN-EE}, and GT-DLA-dsHFF~\cite{GT-DLA-dsHFF}), and two Mamba-based models (U-Mamba~\cite{U-Mamba} and VM-UNet~\cite{VM-UNet}), all implemented using their official codes. SWinMamba outperforms SOTA models on most metrics with low computational cost, achieving average improvements of 3.15\% in $\beta_0$ (reflecting vascular connectivity, the key challenge across all datasets), 0.39\% in clDice, and 0.14\% in Dice across the three datasets. Moreover, visualization (Fig.~\ref{fig3}) demonstrates that SWinMamba achieves superior performance compared to SOTA methods in both completeness and connectivity.

\section{Conclusion}
In this paper, SWinMamba is proposed to achieve complete and connected vascular segmentation for disease diagnosis and treatment. The SWinMamba innovatively models the continuity of slender vascular structures by incorporating serpentine window sequences into bidirectional state space models, adaptively guiding global visual context modeling to the vascular structure. Extensive experiments demonstrate that SWinMamba achieves superior completeness and connectivity with low computational cost and exhibits strong universality. The proposed method has great potential in clinical disease diagnosis and surgical navigation. 

\bibliographystyle{splncs04}
\bibliography{mybibliography}

\end{document}